\documentclass[11pt,a4paper,superscriptaddress]{article}
\usepackage[utf8]{inputenc}

\usepackage{amsmath,amssymb}
\usepackage{color}
\usepackage{xcolor}
\usepackage{graphicx}
\usepackage{subfig}
\usepackage{cite}                
\definecolor{mycitecolor}{RGB}{0,123,167}     
\definecolor{mylinkcolor}{RGB}{33,64,154}     
\definecolor{myurlcolor}{RGB}{134,80,164}     

\usepackage[colorlinks=true,
            linkcolor=mylinkcolor,
            citecolor=mycitecolor,
            urlcolor=myurlcolor]{hyperref}

\usepackage{multirow,makecell}
\usepackage{braket,bm}
\usepackage{physics}
\usepackage{cancel}
\usepackage{leftidx}
\usepackage[text={17cm,24.5cm},centering]{geometry}
\usepackage{textcomp}
\usepackage{wasysym}
\usepackage{mathtools}
\numberwithin{equation}{section}

\def \be {\begin{equation}}
\def \ee {\end{equation}}
\def \ba {\begin{array}}
\def \ea {\end{array}}
\def \bea{\begin{eqnarray}}
\def \eea{\end{eqnarray}}

\def \a {\alpha}

\def \g {\gamma}
\def \G {\Gamma}
\def \d {\delta}

\def \dg {\dagger}
\def \e {\epsilon}

\def \m {\mu}
\def \n {\nu}
\def \k {\kappa}
\def \l {\lambda}

\def \s {\sigma}

\def \th {\theta}

\def \mL {\mathcal L}

\def \mO {\mathcal O}

\def \mS {\mathcal S}
\def \mT {\mathcal T}

\def \mZ {\mathcal Z}

\def \p {\partial}

\def \td {\tilde}

\def \Tr {{\textrm{Tr}}}

\def \and {{\textrm{and}}}

\usepackage[normalem]{ulem}

\begin{document}
\begin{titlepage}
	
	\title{\textbf {Symmetry resolved entanglement entropy after an inhomogeneous quench}}

%
	\author{Hui-Huang Chen,$^{a, b}$\footnote{chenhh@jxnu.edu.cn, corresponding author}~~Xin-Liang Zhou,$^{a}$\footnote{zhouxinliang738@gmail.com}~~Jiayu Yin,$^{a}$\footnote{jiayuyin@jxnu.edu.cn}~~Ming Zhang$^{c,a,d}$ \footnote{mingzhang@jxnu.edu.cn}}

	\date{}
	
	\maketitle
	\underline{}
	\vspace{-12mm}
	
	\begin{center}
		{
          \it
            $^{a}$ College of Physics and Communication Electronics, Jiangxi Normal University,\\ Nanchang 330022, China\\
            $^{b}$ SISSA and INFN Sezione di Trieste, via Bonomea 265, 34136 Trieste, Italy\\
            $^{c}$ Department of Physics and Astronomy, University of Waterloo, Waterloo, Ontario N2L 3G1, Canada\\
            $^{d}$ Perimeter Institute for Theoretical Physics, 31 Caroline St. N., Waterloo, Ontario N2L 2Y5, Canada

		}
		\vspace{10mm}
	\end{center}
\begin{abstract}
We investigate the non-equilibrium time evolution of symmetry-resolved entanglement entropy (SREE) following an inhomogeneous quench in a critical one-dimensional free fermion system. Using conformal field theory, we derive an exact expression for the SREE and analyze its behavior. We find that, at leading order in the long-time limit, the SREE grows logarithmically as 
$\log t$. While the equipartition of entanglement holds at leading order, we identify subleading corrections that break it. Our numerical simulations corroborate the analytical predictions with excellent agreement.
\end{abstract}
	
\end{titlepage}

\thispagestyle{empty}

\newpage

\tableofcontents
\section{Introduction}
The study of non-equilibrium evolution in quantum many-body systems has become a central theme in recent research. During the past decades, researchers realized that entanglement is an essential tool in non-equilibrium quantum physics. In isolated quantum systems, entanglement entropy is a fundamental quantity for characterizing how systems thermalize. The rate of entanglement growth can serve as a measure of quantum chaos and scrambling, relevant in both quantum many-body physics \cite{nahum2017quantum, zhou2020entanglement} and black hole problems \cite{Ryu:2006bv,Takayanagi:2010wp,Almheiri:2020cfm,Chen:2020hmv,Grimaldi:2022suv}.
\par Many real-world systems are not translationally invariant. In cold atom experiments, ultracold gases are trapped using optical lattices, where the trapping potential is typically harmonic rather than uniform. In most quantum systems, introducing inhomogeneity (i.e., spatial dependence in couplings, interactions, or external potentials) breaks integrability and prevents exact solutions. However, certain classes of inhomogeneous Hamiltonians preserve solvability. In quantum many-body systems and conformal field theory  (CFT), sine-square deformation (SSD) \cite{2009PThPh.122..953G,2009PThPh.122..953Gerr,Hikihara:2011mtb,Katsura:2011ss,2011PhRvA..83e2118G,Shibata:2011jup,Maruyama:2011njv,Katsura:2011zyx} and Möbius deformation \cite{Okunishi:2016zat,Wen:2016inm,Goto:2021sqx,Goto:2023wai,Nozaki:2023fkx} are two of the most widely studied inhomogeneous Hamiltonians that preserve solvability. These deformations modify energy density across space while maintaining partial conformal symmetry, leading to interesting applications in condensed matter physics \cite{Wen:2016inm,Wen:2018vux,Goto:2023yxb} and holography \cite{maccormack2019holographic, Goto:2023wai,Bernamonti:2024fgx}. 
\par Conformal field theories in two dimensions are characterized by their infinite-dimensional Virasoro symmetry, generated by holomorphic and antiholomorphic operators $L_n$ and $\bar{L}_n$ ($n \in \mathbb{Z}$). The vacuum state $\ket{0}$ must be invariant under global conformal transformations, which means that it must be annihilated by $L_{-1}, L_0, L_{1}$ and there anti-holomophic counterparts. The sine-square deformation (SSD) and its generalization, the Möbius deformation, modulate the Hamiltonian spatially while retaining ties to conformal symmetry. 
\par As mentioned earlier, in quantum many-body physics, entanglement entropy is a crucial tool for characterizing quantum states. In recent years, symmetry-resolved entanglement entropy (SREE) \cite{Goldstein:2017bua} has garnered significant interest as it provides a finer characterization of entanglement by decomposing the total entanglement entropy into contributions from different symmetry sectors \cite{Murciano:2020vgh, Chen:2021pls, Capizzi:2021zga, Chen:2021nma, Bertini:2022srv, Northe:2023khz,Li:2023zgy, Banerjee:2024ldl,Yan:2024rcl, Huang:2025lsy}. In particular, in nonequilibrium physics, SREE helps us understand how different symmetry sectors evolve independently, which is crucial for studying thermalization and quantum chaos\cite{Feldman:2019upn, Parez:2021pgq, Chen:2022gyy, Oshima:2022yrw, Scopa:2022gfw, Murciano:2023zvk,Caceffo:2024jbc, Chen:2023whs}. For brevity, we reference only key papers here and recommend a recent review \cite{Castro-Alvaredo:2024azg} for a complete reference list. 
\par In nonequilibrium physics, the quantum quench problem in 2D CFT has gained significant attention due to its exact solvability \cite{Das:2017sgp}, universal features \cite{Das:2014jna,Das:2014hqa}, and deep connections to statistical mechanics, holography \cite{Myers:2017sxr}, and quantum information theory. For a detailed discussion of quantum quenches in CFT and its applications, see \cite{calabrese2016quantum} and references therein.
\par Quantum quenches offer a rich framework for exploring non-equilibrium dynamics, with notable examples including global quenches, local quenches, and inhomogeneous quenches. In this paper, we will discuss the quench dynamics of SREE in inhomogeneously deformed CFT. More explicitly, we will consider the following setup: a one-dimensional critical system with open boundary conditions, in the ground state of an uniform Hamiltonian at $t=0$. When $t>0$, we suddenly change the Hamiltonian to be a Möbius/SSD deformed one, and study the evolution of the SREE after this quench. To the best of our knowledge, there are no similar discussions about the dynamics of SREE in such an inhomogeneous quench protocol. Most importantly, our analytical method is also applicable to other inhomogeneous quenches with M\"obius/SSD Hamiltonian governing the post-quench dynamics. Our numerical method has broader validity and applies to any inhomogeneous quenches.  
\section{Möbius/SSD deformation in CFT}\label{sec2}
In this section, we will briefly review some basic facts about M\"obius/SSD deformation in CFT.
\subsection{Möbius and SSD Hamiltonians} 
Let's consider a finite one-dimensional system with length $L$ and open boundary conditions are imposed at its ends $x=0$ and $x=L$. The undeformed Hamiltonian can be written as
\be\label{jfoi329p}
H_0 = \int_0^L dx~h(x)\,,
\ee  
where $h(x)$ is the energy density. The SSD and Möbius deformations introduce position-dependent modulation via an envelope function $f(x)$,
\be 
H_0\rightarrow \int_0^L dx~f(x) h(x)\,.
\ee 
The SSD Hamiltonian $H_1$ is obtained by choosing \(f(x) = 2\sin^2\left(\pi x/L\right)\),
\be\label{jf3892p}
H_1 = \int_0^L 2\sin^2\left(\frac{\pi x}{L}\right) h(x) dx\,.
\ee 
It's convenient to introduce the Möbius deformation, which generalizes SSD with a tunable parameter \(\theta \geq 0\). The Möbius deformed Hamiltonian (or M\"obius Hamiltonian for short) $H_1(\theta)$ corresponds to choosing $f(x)=1-\tanh(2\theta)\cos(2\pi x/L)$, and we can rewrite it as 
\be
H_1(\theta) = H_0 - \frac{\tanh(2\theta)}{2} \left(H_+ + H_-\right)\,,
\ee  
where we have defined $H_\pm = \int_0^L e^{\pm 2\pi ix/L} h(x) dx$. For $\theta \to \infty$, the M\"obius Hamiltonian $H_1(\th)$ reduces to the SSD Hamiltonian $H_1$ in \eqref{jf3892p}, while $\th = 0$ recovers the undeformed Hamiltonian $H_0$ in \eqref{jfoi329p}. The M\"obius deformation interpolates between these limits, scaling the effective system size to \(L_{\text{eff}} = L\cosh(2\theta)\). The SSD suppresses energy density at boundaries $x=0, L$, effectively removes the boundary effects of a finite system. 
\par In terms of Virasoro generators, we have 
\be
H_1(\theta) = N_{\theta}^{-1}(\mL_0+\bar\mL_0)\,,
\ee 
where
\begin{align} 
N_{\theta} &=\cosh(2\theta)\,,\\
\mL_0 &=\cosh(2\theta)L_0 - \sinh(2\theta)\frac{L_1 + L_{-1}}{2}\,.
\end{align}
 Remarkably, the SSD ground state coincides with the uniform CFT vacuum under periodic boundary conditions, as $(L_1 + L_{-1})/2$ annihilates the vacuum.  
\par In \cite{Okunishi:2016zat}, the author suggests that Möbius deformed CFT can most conveniently be viewed as CFT quantized on the so-called Möbius coordinate. The procedure is found to be equivalent to consider the following Möbius transformation
\be\label{ztozeta}
z \mapsto \zeta=-\frac{z\,\cosh\theta   - \sinh\theta}{z\,\sinh\theta  - \cosh\theta}
\ee
and define new Virasoro generators in terms of the mapped coordinate
\be 
\mL_n=\oint\frac{d\zeta}{2\pi i}\zeta^{n+1}T(\zeta)\,,
\ee
where
\be 
T(\zeta)=\left(\frac{dz}{d\zeta}\right)^2T(z)=(\cosh\th-z\sinh\th)^4T(z)\,.
\ee
Here we have used the fact that the Schwartz derivative of any M\"obius map is zero.
\par These new Virasoro generators $\mL_n$ satisfy the same Virasoro algebra as the radial quantized CFT. However, at the SSD point, i.e. when $\theta\rightarrow\infty$, and if we define the scaled generator and index as $\mL/N_{\theta}\rightarrow\td{\mL}_n, \, n/N_{\theta}\rightarrow\kappa$, we obtain a continuous Virasoro algebra   
\be 
[\td{\mL}_\k, \td{\mL}_{\k'}] = (\k - \k') \td{\mL}_{\k+\k'} + \frac{c}{12} \k^3 \delta(\k + \k')\,,
\ee    
where $c$ is the central charge and $\k$ becomes a continuous index as $\th\to\infty$. Consequently, SSD CFTs exhibit a continuous energy spectrum, mimicking an infinite system despite finite $L$.  
\section{SREE in free Dirac fermion CFT}
In this section, we will review the fluxed twist field method of computing SREE in free Dirac fermion CFT. Let's first introduce the concept of symmetry-resolved entanglement entropy. In the next subsection, we will discuss the fluxed twist field method.
\subsection{Symmetry-resolved entanglement entropy}
SREE quantifies the entanglement contribution from individual symmetry sectors of a quantum system. For a system with a conserved charge $Q$, the reduced density matrix $\rho_A$ decomposes into blocks labeled by eigenvalues $q$ of $Q_A$, the charge restricted to subsystem $A$. The SREE for sector $q$ is defined as 
\be
S_A(q) = -\Tr\left[\rho_A(q)\log\rho_A(q)\right]\,, \quad \rho_A(q) = \frac{\Pi_q \rho_A \Pi_q}{\text{Tr}(\Pi_q \rho_A)}\,,  
\ee
where \( \Pi_q \) projects onto the $q$-sector. 
\par To compute the SREE, we first introduce the charged moments
\be
Z_n(\alpha) = \Tr\left(\rho_A^n e^{i\alpha Q_A}\right)\,.
\ee   
Then the symmetry-resolved R\'enyi entropies $S_A^{(n)}(q)$ are derived via Fourier transform, 
\be\label{SRRE}
\mZ_n(q) = \int_{-\pi}^\pi \frac{d\alpha}{2\pi} e^{-i\alpha q} Z_n(\alpha)\,, \quad S_A^{(n)}(q) = \frac{1}{1-n}\log\left[\frac{\mZ_n(q)}{\mZ_1(q)^n}\right]\,.  
\ee  
The SSRE $S_A(q)$ is obtained from $S_A^{(n)}(q)$ by taking $n\to 1$ limit. A direct computation leads to 
\be\label{Sq}
S_A(q)=\log\mZ_1(q)-\frac{1}{\mZ_1(q)}\frac{\p\mZ_n(q)}{\p n}\Big|_{n=1}\,,
\ee
where $\mZ_1(q)$ can be interpreted as the physical probability distribution of the charge $Q_A$.
\subsection{Fluxed twist fields method}
In this paper, we will focus on the case where the post quench dynamics is governed by the Möbius/SSD Hamiltonian of a 2D massless free Dirac fermion CFT. 
The action of the massless free Dirac fermion is
\be 
\mS=\int dtdx\bar\psi i\g^{\m}\p_\m\psi\,,
\ee 
where $\bar\psi=\psi^{\dg}\g^0$ and we choose $\g^0=\s_1,\g^1=\s_2$. This theory has a global $U(1)$ symmetry: $\psi\to e^{i\a}\psi, \bar\psi\to e^{-i\a}\bar\psi$. The corresponding conserved charge is $Q=\int dx \psi^{\dg}\psi$. 
\par In CFT, SREE is computed using fluxed twist fields $\mathcal{T}_{n,\m}$, which unify the role of twist operator and flux insertion. The conventional twist field creates an $n$-sheeted Riemann surface for Rényi entropy, and a flux insertion introduces a phase factor \( e^{i\m Q_A} \) to resolve symmetry sectors.  
\par The charged moments are expressed as
\be
Z_n(\alpha) = \Tr\left(\rho_A^n e^{i\alpha Q_A}\right) = \left\langle \mT_{n,\a}(u)\td{\mT}_{n,\a}(v) \right\rangle \,,  
\ee 
where $u, v$ mark the endpoints of subsystem $A = [u, v]$. 
\par In the free Dirac fermion CFT, we can compute SREE using the fluxed twist field method. This method leverages the replica trick and the bosonization technique to rewrite the computation in terms of correlation functions of special vertex operators. Below, we outline the key steps in this approach.
\par To compute the charged moments, we use the replica trick, which reformulates the computation as a partition function on an fluxed $n$-sheeted Riemann surface. Instead of working directly on this complex topology, we equivalently introduce $n$ replicated copies of the Dirac field 
\be 
\Psi = \begin{pmatrix} \psi_1 \\ \psi_2 \\ \vdots \\ \psi_n \end{pmatrix}\,,
\ee
where each $\psi_j$ corresponds to a field living on the $j$-th replica sheet.
\par The key observation is that the topology of the fluxed Riemann surface can be encoded in the boundary conditions imposed at the entangling points (i.e., the edges of the subsystem). These boundary conditions are enforced by inserting fluxed twist field operators $\mT_{n,\a}$ and fluxed anti-twist field operators $\td{\mT}_{n,\a}$, which cyclically permute the replica fields while adding a phase $e^{i\a/n}$ related to the flux.
\par When the field $\psi_j$ is transported around the twist field insertion at $u$, it transforms as
\be\label{T}
\psi_j \to e^{i\a/n} \psi_{j+1}\,,
\ee
while the fluxed anti-twist field $\td{\mT}_{n,\a}$ acts as the inverse transformation
\be
\psi_j \to e^{-i\a/n} \psi_{j-1}\,.
\ee
The transformation \eqref{T} is encoded in a twist matrix
\be
T_\a = \begin{pmatrix} 
0 & -e^{i\a/n} & 0 & \cdots & 0 \\
0 & 0 & -e^{i\a/n} & \cdots & 0 \\
\vdots & \vdots & \vdots & \ddots & \vdots \\
e^{i\a/n} & 0 & 0 & \cdots & 0
\end{pmatrix}\,.
\ee
To decouple the replicas, we diagonalize this transformation by defining a new basis of fields
\be
\td{\psi}_k = \sum_{j=1}^{n} e^{-i(2\pi k j / n - \pi j)} \psi_j, \quad k = -\frac{n-1}{2}\,, \dots, \frac{n-1}{2}\,,
\ee
where the new fields $\td{\psi}_k$ transform independently under the twist operator
\be
T_{\a} \td{\psi}_k = e^{i(\a + 2\pi k)/n} \td{\psi}_k\,.
\ee
Since each mode $\td{\psi}_k$ is now independent, $\mT_{n,\a}$ and $\td{\mT}_{n,\a}$ can be decomposed into $n$ different twists
acting independently on each replica
\be\label{fluxedtwist}
\mT_{n,\a}=\prod_{k=-\frac{n-1}{2}}^{\frac{n-1}{2}}\mT_{n,k,\a}\,,\qquad \td{\mT}_{n,\a}=\prod_{k=-\frac{n-1}{2}}^{\frac{n-1}{2}}\td{\mT}_{n,k,\a}\,.
\ee
The partition function also factorizes, allowing us to express the charged moments as a product
\be\label{Znalpha}
Z_n(\a) = \prod_{k=-\frac{n-1}{2}}^{\frac{n-1}{2}} \left\langle \mT_{n,k,\a}(u)\td{\mT}_{n,k,\a}(v) \right\rangle\equiv\prod_kZ_k(\a)\,,
\ee
where each factor $Z_k(\a)$ corresponds to a correlation function of twist fields for a single decoupled mode.
\par As shown above, by diagonalizing the matrix $T_{\a}$ through a unitary transformation, we obtain $n$ decoupled fields $\td{\psi}_k$, each associated with an eigenvalue $e^{i(\a + 2\pi k)/n}$. These fields satisfy the twisted boundary conditions
\be 
\td{\psi}_k(e^{2\pi i}u)= e^{i(\a/2\pi+k)/n} \td{\psi}_k(u)\,, \qquad \td{\psi}_k(e^{2\pi i}v)=e^{-i(\a/2\pi+k)/n}\td{\psi}_k(v)\,. 
\ee
Instead of dealing with these multivalued fields directly, we introduce an external gauge field $A_\m^k$, which enforces the same boundary conditions while allowing $\td{\psi}_k(x)$ to remain single-valued. Thus the Lagrangian density is
\be
\mL_k = \bar{\td{\psi}}_k \g^\m (\p_\m + i A_\m^k) \td{\psi}_k\,.
\ee
The gauge field \( A_\m^k \) is a pure gauge everywhere except at the points $u$ and $v$, where it has a vortex-like singularity. The flux through any contour surrounding a branch point is given by
\be
\oint_{C_{u}} dx^\m A^k_\m = -\frac{2\pi k+\a}{n}\,, \quad \oint_{C_{v}} dx^\m A^k_\m = \frac{2\pi k+\a}{n}\,.
\ee
These conditions imply that \( A_\mu^k \) satisfies the equation
\be\label{A} 
\epsilon^{\m\n} \p_\n A_\m^k(x) = \frac{2\pi k+\a}{n}\left[ \delta(x - u) - \delta(x - v) \right]\,.
\ee
With these transformations, the original partition function factorizes into a product over independent modes
\be
Z_n(\a)=\prod_{k=-\frac{n-1}{2}}^{\frac{n-1}{2}} Z_k(\a)\,.
\ee
Each $Z_k$ can now be rewritten in terms of a vacuum expectation value in the presence of the external gauge field as
\be\label{Zk}
Z_k(\a)= \left\langle e^{i \int A_\mu^k j^\mu_k d^2x} \right\rangle \,.
\ee
\subsection{Bosonization}  
Bosonization is a powerful technique in 1+1D quantum field theory that maps fermionic systems to bosonic fields, making certain calculations more tractable. In the case of free Dirac fermions, the bosonization procedure allows us to express charged moments $Z_n(\a)$ in terms of correlation functions of vertex operators in a free bosonic theory. The fermion current $j_k^\m$, which generates the global $U(1)$ symmetry associated with charge conservation, is mapped to the bosonic field as
\be
j_k^\mu = \bar{\td{\psi}}_k \g^\m\td{\psi}_k \quad \longrightarrow \quad \frac{1}{2\pi} \epsilon^{\mu\nu} \p_\nu \phi_k \,.
\ee
The free Dirac fermion $\td{\psi}_k$ is mapped to a free compact boson field $\phi_k(x)$ with the action
\be
S = \frac{1}{8\pi} \int d^2x \, (\p_\m\phi_k)^2 \,.
\ee
Then using Eq.~(\ref{A}) and integrating by part, the right-hand-side of Eq.~(\ref{Zk}) becomes correlation function of vertex operators
\be
Z_k(\a) = \left\langle e^{i\frac{k+\alpha/2\pi}{n}\phi_k(u)} e^{-i\frac{k+\alpha/2\pi}{n}\phi_k(v)} \right\rangle \,.
\ee 
Comparing the equation above with Eq.~(\ref{Znalpha}), one can identify 
\be
\mT_{n,k,\a}=e^{i\frac{k+\alpha/2\pi}{n}\phi_k}=V_{a,a}\,,\qquad 
\td{\mT}_{n,k,\a}=e^{-i\frac{k+\alpha/2\pi}{n}\phi_k}=V_{-a,-a}\,.
\ee
Recall that the conformal dimensions of vertex operator $V_{a,\bar a}\equiv e^{ia\varphi+i\bar a\bar\varphi}$ are $(h,\bar h)=(\frac12 a^2,\frac12\bar a^2)$. Here $\varphi,\bar\varphi$ are the holomorphic and anti-holomorphic part of the scalar field $\phi$. Therefore, the conformal dimensions of the fluxed twist fields in Eq.~(\ref{fluxedtwist}) are the same,
\be 
h_{n,\a}=\bar h_{n,\a}=\frac{1}{2}\sum_{k=-\frac{n-1}{2}}^{\frac{n-1}{2}}\left(\frac{k}{n}+\frac{\a}{2\pi n}\right)^2
=\frac{1}{24}\left(n-\frac{1}{n}\right)+\frac{\a^2}{8\pi^2 n}\,.
\ee
\section{Inhomogeneous quench dynamics from fluxed twist fields}  
In this section, we will compute the time evolution of SREE after a quench from uniform to non-uniform CFT \cite{Wen:2018vux}. Specificly, we will consider a critical free-fermion chain of length $L$ with open boundary conditions, initially prepared in its ground state, i.e. the ground state of the free Dirac fermion CFT (the uniform one). At $t=0$, we evolve the system according to the Hamiltonian of the sine-square deformed CFT, which is certainly a non-uniform one.  
\subsection{Time evolution under SSD Hamiltonian}
The initial state is the ground state $\ket{\psi_0}$ of a uniform CFT, defined on a finite spatial interval $[0, L]$ with open boundary conditions. The subsystem $A$ is $[0,l]$. We choose open boundaries because, as highlighted in Section \ref{sec2}, a periodic boundary condition yields a ground state identical to that of the SSD system, leading trivial dynamics. At time $t = 0$, we abruptly switch the Hamiltonian to the M\"obius Hamiltonian $H_{1}(\theta)$. In the Schr\"odinger picture, the state evolves as $\ket{\psi(t)}= e^{-i H_{1}(\theta) t}\ket{\psi_0}$ and the correlation functions evolves as $\bra{\psi_0} e^{i H_{1} t} O(x_1) \cdots O(x_n) e^{-i H_{1} t}\ket{\psi_0}$, where $O(x_i)$ are operators inserted at spatial points $x_i$. Then the charged moments of the reduced density matrix $\rho_A$ at time $t$ are  
\be\label{Znat1}
Z_n(\a, t) = \bra{\psi_0}e^{iH_{1}(\th)t} \mathcal{T}_{n,\a}(w_0,\bar{w}_0) e^{-iH_{1}(\th)t}\ket{\psi_0}\,.
\ee 
The system has conformal boundary conditions imposed at $\sigma = 0$ and $\sigma = L$, which are assumed to be identical here for simplicity.
Here we introduce the complex coordinate $w = \tau + i\sigma$ where $-\infty < \tau < \infty$ and $0 \leq \sigma \leq L$, and $w_0 = il$ marks the fluxed twist operator’s position. 
\par It's convenient to shift to Euclidean spacetime by setting $it=\tau$, transforming the charged moments Eq.~(\ref{Znat1}) into
\be 
Z_n(\a,\tau)=\bra{\psi_0}e^{H_{1}(\theta) \tau} \mathcal{T}_{n,\a}(w_0,\bar{w}_0) e^{-H_{1}(\theta) \tau}\ket{\psi_0}\,.
\ee
Rather than evolving the state in the Schr\"odinger picture, we turn to the Heisenberg picture and evolve the operator $\mathcal{T}_{n,\a}$ under $H_{1}(\theta)$. It turns out that a suitable conformal transformation simplifies this evolution, making it tractable. The M\"obius Hamiltonian $H_{1}(\theta)$ can be viewed as a regularized version of $H_{1}$ that avoids potential infrared (IR) divergences tied to the SSD’s continuous spectrum. We first consider the Möbius Hamiltonian $H_{1}(\theta)$, then take the limit $\theta \to \infty$ to recover the SSD case. 
\par The Möbius Hamiltonian in the $w$-plane, expressed via the stress-energy tensor, is
\be 
H_{1}(\theta) = H_0 - \frac{\tanh (2 \theta)}{2} (H_+ + H_-)\,,
\ee
where
\be 
H_0 = \int_0^L \frac{d \sigma}{2 \pi} T_{\tau \tau}(\sigma) = \int_0^L \frac{d \sigma}{2 \pi} [T(w) + \bar{T}(\bar{w})]\,, 
\ee
and
\be 
H_{\pm} = \int_0^L \frac{d \sigma}{2 \pi} \left[ e^{\pm 2 \pi w / L} T(w) + e^{\mp 2 \pi \bar{w} / L} \bar{T}(\bar{w}) \right]\,. 
\ee
In the limit $\theta\to\infty$, this reduces to the SSD Hamiltonian $H_1 = H_0 - \frac{1}{2} (H_+ + H_-)$.
\par Then we map the strip in the $w$-plane to the whole complex plane via $z=e^{2\pi w/L}$. In the $z$-plane, the Möbius Hamiltonian $H_1(\theta)$ becomes
\be 
\begin{split}
H_1^{(z)}(\theta)=&\frac{2\pi}{L\cosh(2\theta)} \left[\cosh(2\theta)\oint\frac{zdz}{2\pi i}T(z)-\frac{\sinh(2\theta)}{2}\oint \frac{z^2 + 1}{2\pi i}T(z)dz \right]\\
&+\frac{2\pi}{L\cosh(2\theta)} \left[\cosh(2\theta)\oint\frac{\bar zd\bar z}{2\pi i}\bar T(\bar z)-\frac{\sinh(2\theta)}{2}\oint \frac{\bar z^2 + 1}{2\pi i}\bar T(\bar z)d\bar z \right]-\frac{\pi c}{6L}\,,
\end{split}
\ee
where $c=1$ for Dirac fermion CFT. This form remains complicated, so we apply a further Möbius transformation to the $\zeta$-plane (see also Eq.~(\ref{ztozeta})
\be\label{zeta}
\zeta(z) = -\frac{z\cosh\th-\sinh\th}{z\sinh\th -\cosh\th}\,,
\ee
simplifying the Hamiltonian to
\be 
H_{1}^{(\zeta)}(\theta) = \frac{2\pi}{L\cosh(2\theta)}(L_0^{(\zeta)}+\bar L_0^{(\zeta)})-\frac{\pi c}{6L}\,, 
\ee
where $L_0^{(\zeta)}$ is the zero-mode Virasoro generator in the $\zeta$-plane. From the equation above, it follows that in the $\zeta$-plane, the time evolution acts as a dilatation,
\be 
e^{H_1^{(\zeta)}(\th) \tau} \mathcal{T}_{n,\a}^{(\zeta)}(\zeta, \bar{\zeta}) e^{-H_1^{(\zeta)}(\th) \tau} = \l^{2 h_{n,\a}} \mathcal{T}_{n,\a}^{(\zeta)}(\l\zeta, \l \bar{\zeta})\,, 
\ee
with $\l=\exp\left(\frac{2\pi\tau}{L\cosh(2\theta)}\right)$. 
Mapping back to the $z$-plane, this corresponds to shift the location of $\mT_{n,\a}^{(z)}$ from $z$ to $z'$, where $z'$ is determined through the equation $\zeta(z') = \l \zeta(z)$. Solving this equation, one obtains
\be\label{zp}
z'=\frac{[(1-\l)\cosh(2\theta)-(\l + 1)]z +(\l-1)\sinh(2\theta)}{(1-\l)\sinh(2\theta)z + [(\l-1)\cosh(2\theta)-(\l + 1)]}\,. 
\ee
\par After a series of conformal mappings described above, the computation of the charged moments are equivalent to evaluate the one-point function of the fluxed twist field on the $z$-plane
\be 
\langle\psi_0| e^{H_1(\theta) \tau} \mathcal{T}_{n,\a}^{(w)}(w_0, \bar{w}_0) e^{-H_1(\theta) \tau} |\psi_0\rangle = \left(\frac{\p z}{\p w}\frac{\p\bar z}{\p\bar w}\frac{\p z'}{\p z}\frac{\p\bar z'}{\p\bar z}\Big|_{w=w_0}\right)^{h_{n,\a}}\langle\mathcal{T}_{n,\a}^{(z)}(z',\bar z')\rangle\,.
\ee
Then we need to compute the one-point function $\langle\mathcal{T}_{n,\a}^{(z)}(z,\bar z) \rangle$ in the $z$-plane with a slit along the real axis from $[0, \infty)$. A further map $z\to z^{1/2}$ map this plane to the upper half plane with conformal boundary condition imposed on the real axies. Then the image method of boundary conformal field theory can be applied and the finial result is
\be 
\langle \mathcal{T}_{n,\a}^{(z)}(z, \bar z)\rangle = \left(\frac14 z^{-\frac12} \bar z^{-\frac12} \right)^{h_{n,\a}}\left(\frac{2i\e }{z^{\frac12}-\bar z^{\frac12}}\right)^{2h_{n,\a}}\,, 
\ee
where $\e$ is a UV cutoff (e.g., lattice spacing). In our case, the UV cutoff $\e$ actually depends on the R\'enyi index $n$ and the flux $\a$, and can be fixed by comparing with previous known results in the equilibrium case. 
\subsection{The charged moments}\label{sec4.2}
\par The explicit form of the charged moments in the M\"obius case is very complicated and we do not report the final result here. However, taking $\th \to \infty$ and performing the analytic continuation $\tau \to it$, we can obtain a result with a relatively concise expression for the time-dependent charged moments of SSD quench
\be 
Z_n(\a,t) = \left(\frac{2 \pi^2 \e^2}{L^2} \right)^{h_{n,\a}}\left(r(t)^2 + r(t)s(t)\right)^{-h_{n,\a}}\,,
\ee
where we have defined the functions
\be 
r(t) = \sqrt{s(t)^2 + \sin^2 \frac{2 \pi l}{L}}\,,
\ee
\be 
s(t) = \frac{4\pi^2 t^2}{L^2}\sin^2\frac{\pi l}{L}-\cos\frac{2\pi l}{L}\,.
\ee
At $t=0$, we have $r(0)=1,s(0)=-\cos\frac{2\pi l}{L}$ and thus
\be\label{Znat0} 
\log Z_n(\a,t=0)=-h_{n,\a}\log\left[\frac{L^2}{2\pi^2\e^2}\left(1-\cos\frac{2\pi l}{L}\right)\right]\,.
\ee
This should equal to the equilibrium result of the charged moments \cite{Bonsignori:2020laa}
\be\label{Zna0}
\log Z_n(\a)=-\left[\frac{1}{12}\left(n-\frac{1}{n}\right)+\frac{1}{n}\left(\frac{\a}{2\pi}\right)^2\right]\log\left[\frac{4L}{\pi}\sin\left(\frac{\pi l}{L}\right)\right]+\frac{\Upsilon(n,\a)}{2}\,,
\ee
where $\Upsilon(n,\a)$ is a real constant and is determined by exploiting the Fisher-Hartwig conjecture
\be 
\Upsilon(n,\a)=ni\int_{-\infty}^{\infty}dw\log \frac{\G\left(\frac{1}{2}+i w\right)}{\G\left(\frac{1}{2}-i w\right)} \left(\tanh(\pi w)-\tanh\left(\pi n w+\frac{i\a}{2}\right)\right)\,.
\ee
By comparing Eq.~(\ref{Znat0}) and Eq.~(\ref{Zna0}), we can eliminate the UV cut-off $\e$, and the finial result of time dependent charged moments is\footnote{The UV cut-off $\e$ is fixed by the equation $4h_{n,\a}\log(4\e)=\Upsilon(n,\a)$.}
\be\label{Znat}
\log Z_n(\a,t)=-h_{n,\a}\log\left[\frac{8L^2}{\pi^2}\left(r(t)^2+r(t)s(t)\right)\right]+\frac{\Upsilon(n,\a)}{2}\,.
\ee 
\par In \cite{Bonsignori:2019naz}, it was argued that the $\a$-dependence in the cutoff $\e$ can be very well described by keep up to the $\a^2$ term in the expansion of $\e$ at $\a=0$.
Therefore we expand $\Upsilon_n(\a)$ up to $\a^2$ as
\be
\Upsilon_n(\a)=\Upsilon_n(0)+\g(n)\a^2+\mO(\a^4)\,,
\ee
with 
\be 
\g(n)=\frac12\frac{\p^2\Upsilon_n}{\p\a^2}\Big|_{\a=0}=\frac{ni}{4}\int_{-\infty}^{\infty}dw\log \frac{\G\left(\frac{1}{2}+i w\right)}{\G\left(\frac{1}{2}-i w\right)} \left(\tanh^3(n\pi w)-\tanh\left(\pi n w\right)\right)\,.
\ee
Then we can approximate $\log Z_n(\a,t)$ by expanding in $\a$ up to the quadratic term as
\be 
\log Z_n(\a,t)\simeq \log Z_n(0,t)-\frac{\a^2}{2}b_n(t)\,,
\ee
or equivalently 
\be 
Z_n(\a,t)\simeq Z_n(0,t)e^{-\frac{\a^2}{2}b_n(t)}\,.
\ee
Then $Z_n(\a,t)$ is a Guassian in $\a$ with a time-dependent variance $1/b_n(t)$. Here we have introduce the function $b_n(t)$ as
\be 
b_n(t)=\frac{1}{4n\pi^2}\log\left[\frac{8L^2}{\pi^2}(r(t)^2 + r(t)s(t))\right]-\g(n)\,.
\ee
For later use, we also define a related function $\mathfrak{b}(t)$ as\footnote{$\g'(1)\equiv\p_n\g(n)|_{n=1}=\g(1)+\frac{i}{4}\int_{-\infty}^{\infty}dw~\pi w (\cosh(2\pi w)-2)\text{sech}^4(\pi w)\log \frac{\G\left(\frac{1}{2}+i w\right)}{\G\left(\frac{1}{2}-iw\right)}.$}
\be 
\mathfrak{b}(t)\equiv\frac{\p b_n(t)}{\p n}\Big|_{n=1}=-\frac{1}{4\pi^2}\log\left[\frac{8L^2}{\pi^2}(r(t)^2 + r(t)s(t))\right]-\g'(1)\,.
\ee
The total R\'enyi entropies can be obtained from the charged moments at $\a=0$ as
\be 
S_A^{(n)}(t)=\frac{1}{1-n}\log Z_n(0,t)=\frac{n+1}{24n}\log\left[\frac{8L^2}{\pi^2}\left(r(t)^2+r(t)s(t)\right)\right]+\frac{\Upsilon(n)}{2(1-n)}\,,
\ee
where $\Upsilon(n)\equiv\Upsilon(n,0)$ is the non
universal constant obtained in \cite{jin2004quantum} in the absence of fluxes. The total von Neumann entropy is obtained from $S_A^{(n)}(t)$ by taking the limit $n\to 1$ as\footnote{$\Upsilon'(1)\equiv\p_n\Upsilon(n)|_{n=1}=-\pi i\int_{-\infty}^{\infty}dw~w\text{sech}^2(\pi w)\log \frac{\G\left(\frac{1}{2}+i w\right)}{\G\left(\frac{1}{2}-iw\right)}.$} 
\be\label{SvN}
S_A(t)=\frac{1}{12}\log\left[\frac{8L^2}{\pi^2}\left(r(t)^2+r(t)s(t)\right)\right]-\frac{\Upsilon'(1)}{2}\,.
\ee 
\subsection{Symmetry resolved entanglement entropy}
We now turn to the computation of the symmetry-resolved entanglement from the charged moments obtained in the previous section (c.f. Eq.~(\ref{SRRE})). 
$\mZ_n(q, t)$ are obtained by Fourier transforming $Z_n(\a, t)$,  
\be
\mZ_n(q, t) = \int_{-\pi}^\pi \frac{d\a}{2\pi} e^{-i\a q} Z_n(\a, t)\,.
\ee  
Substituting \(Z_n(\alpha, t)\), the integral becomes Gaussian in $\a$. The result is
\be\label{Znq0}
\mZ_n(q, t)=Z_n(0,t)\frac{e^{-\frac{q^2}{2b_n}}}{\sqrt{2\pi b_n}}\left[\erf\left(\frac{b_n\pi+iq}{\sqrt{2b_n}}\right)+\erf\left(\frac{b_n\pi-iq}{\sqrt{2b_n}}\right)\right]\,,
\ee  
where $\erf(x)$ is the error function
\be\label{erf}
\erf(x)=\frac{2}{\sqrt{\pi}}\int_0^{x}dte^{-t^2}\xrightarrow{x\rightarrow\infty}1-\frac{e^{-x^2}}{\sqrt{\pi}x}\,.
\ee
In the large $L$ limit, $b_n$ scale as $\log L$, taking into account that $q\in\mathbb{Z}$, at leading order, we have\footnote{If one using the asymptotic expansion of the error function in Eq.~(\ref{erf}) to evaluate Eq.~(\ref{Znq0}), one will obtain $\mZ_n(q, t)\simeq\frac{Z_n(0,t)}{\sqrt{2\pi b_n(t)}}e^{-\frac{q^2}{2 b_n(t)}}-Z_n(0,t)\frac{b_n(t)e^{-\frac{1}{2}\pi ^2 b_n(t)} (-1)^q}{\pi^2 b_n(t)^2+q^2}$. However, the second term is exponentially suppressed in the large $L$ limit, which can be omitted.}  
\be 
\mZ_n(q, t)\simeq\frac{Z_n(0,t)}{\sqrt{2\pi b_n(t)}}e^{-\frac{q^2}{2 b_n(t)}}\,.
\ee  
The symmetry resolved Rényi entropy in sector $q$ is 
\be
S_A^{(n)}(q, t) = S_A^{(n)}(t)-\frac12\log(2\pi)+\frac{1}{1-n}\log\frac{b_1(t)^{n/2}}{b_n(t)^{1/2}}-\frac{q^2}{2(1-n)}\left(\frac{1}{b_n(t)}-\frac{n}{b_1(t)}\right)\,.
\ee
Taking the limit $n\to 1$ of $S_A^{(n)}(q,t)$, we obtain the time-dependent SREE   as  
\be 
S_A(q, t) =S_A(t)-\frac{1}{2}\log(2\pi b_1(t))+\frac{\mathfrak{b}(t)}{2 b_1(t)}-\frac{\mathfrak{b}(t)+b_1(t)}{2 b_1(t)^2}q^2\,.
\ee 
The leading term is the total entanglement entropy, which scales as $\log L$. The next two terms are order $\log(\log L)$ and order $L^0$ respectively and both do not depend on $q$. Up to this order, the equipartition of entanglement \cite{Xavier:2018kqb} is valid. The first term breaks the equipartition of entanglement aprears at order $\frac{1}{\log L}$. 
\par \begin{figure}
        \centering
        \subfloat
        {\includegraphics[width=7cm]{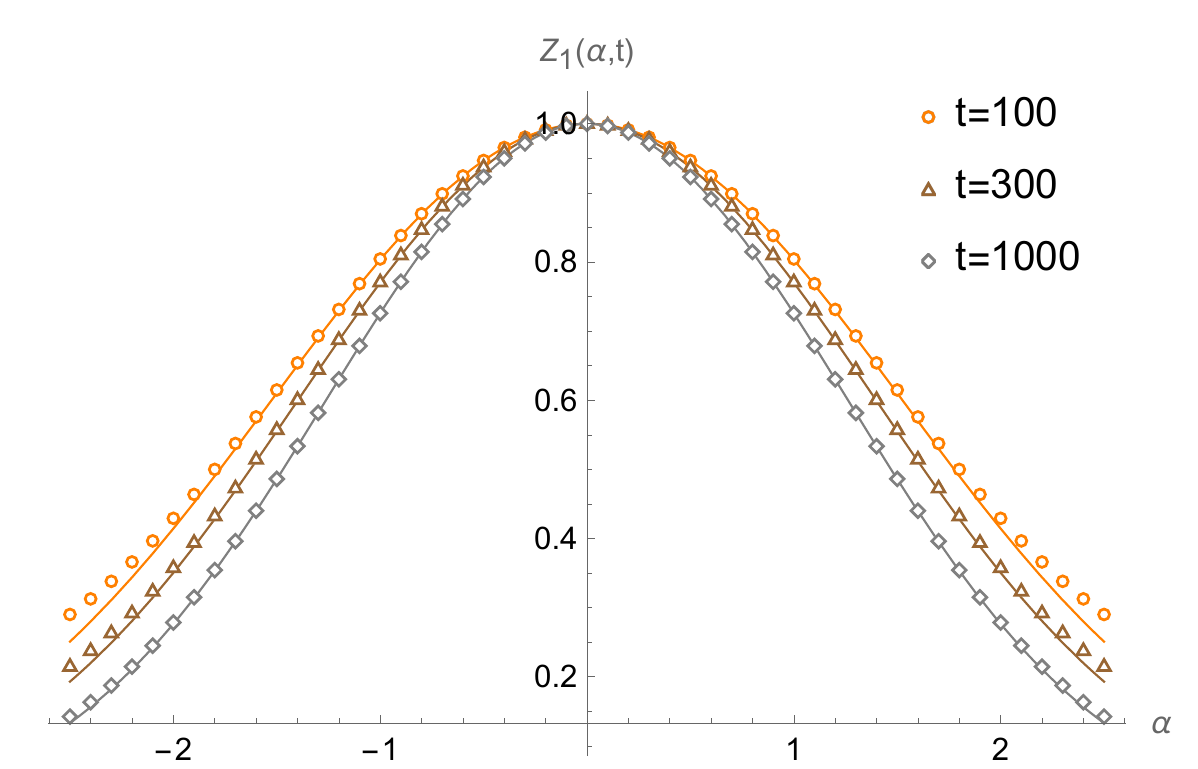}}\qquad\qquad
        {\includegraphics[width=7cm]{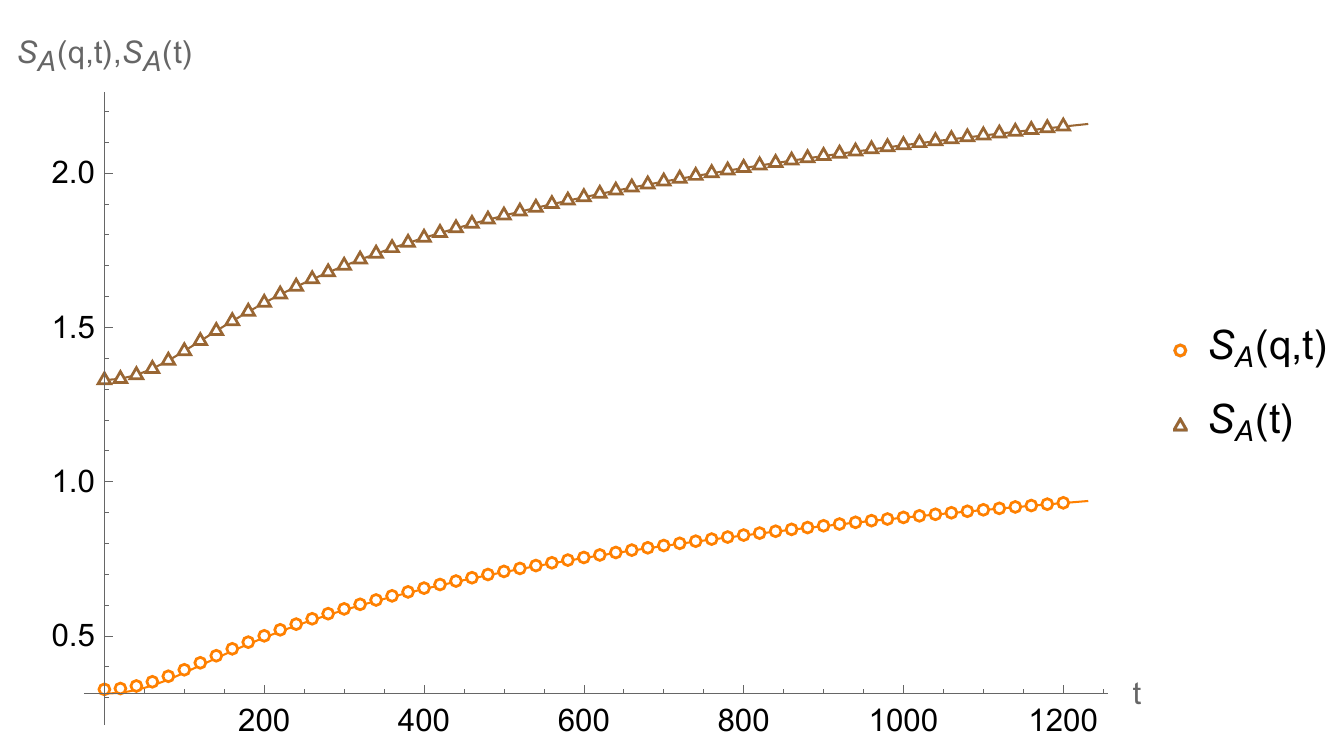}} 
        \caption{The numerical test of our analytical predictions. The full lines are the analytical results. We take $L=600$ and $l=200$. Left panel: the charged moment $Z_1(\a,t)$ as a function of $\a$ at different time. Right panel: The SREE $S(q,t)$ as a function of $t$, here we take $q=1$. As a comparison, we also plot the time evolution of the von Neumann entropy $S_A(t)$ in the same figure. As shown in the figure, the agreement is perfect.}
        \label{fig1}
\end{figure}
\subsection{Late time behavior}
The right panel of Fig.~\ref{fig1} illustrates the time evolution of the SREE and the von Neumann entropy. Notably, these quantities exhibit no revivals despite the finite size of the system. This behavior contrasts with the uniform case, where entanglement revivals arise from quasiparticle reflections at the boundaries. The absence of revivals suggests that sine-square deformation effectively suppresses boundary effects in finite systems. A quasiparticle picture provides insight into this phenomenon, at least for the von Neumann entropy \cite{Wen:2018vux}. Specifically, the local group velocity of quasiparticles, given by $v(x) = 2\sin^2\frac{\pi x}{L}$, vanishes as $x \to 0, L$, implying an infinitely long reflection time at the boundaries. Given that the long-time behavior of the SREE is governed by the leading-order contribution of the von Neumann entropy, the suppression of revivals is expected to extend to the SREE as well.

We next analyze the late-time dynamics of the SREE. In the long-time limit $ t \gg 1$, the quantities $r(t)$ and $s(t)$ asymptotically approach $t^2$, while $b_n(t) \sim \log t $. From Eq.~\eqref{SvN}, the total entanglement entropy thus grows logarithmically as $S_A(t) \sim \log t $, unbounded as $ t \to \infty$. The leading-order behavior of the SREE similarly exhibits logarithmic growth in this regime. However, the SREE incorporates subleading corrections, which become $q$-dependent at $\mathcal{O}\left(\frac{1}{\log t}\right)$. This distinction highlights the nuanced role of symmetry resolution in entanglement dynamics at subleading orders.
\section{Numerical benchmark}
In this part, we will check our previous analytical predictions against the numerical method. We consider the free fermion on a 1d lattice with length $L$. We assume an open boundary condition is imposed. We initially prepare the system in the ground state of the uniform Hamiltonian
\be\label{H0} 
H_0=\frac12\sum_{n=1}^{L-1}c_n^{\dg}c_{n+1}+c_{n+1}^{\dg}c_n\,,
\ee 
where $c_n$ satisfy the canonical anti-commuting relation $\{c_m,c_n^{\dg}\}=\d_{mn}$. Since this uniform Hamiltonian is translation invariant, we can diagonalize it by discrete Fourier transform
\be 
d_k=\sqrt{\frac{2}{L+1}}\sum_{n=1}^{L}\sin\left(\frac{\pi kn}{L+1}\right)c_n\,,\qquad \{d_k,d_{k'}^{\dg}\}=\d_{kk'}\,.
\ee
In terms of this new Bogoliubov fermions, the Hamiltonian (\ref{H0}) has the diagonal form
\be 
H_0=\sum_{k=1}^{L}\e(k)d_k^{\dg}d_k
\ee
with the dispersion relation $\e(k)=\cos\left(\frac{\pi k}{L+1}\right)$. The ground state is the state that all negative energy levels are occupied, which means
\be 
\langle d_k^{\dg}d_{k'}\rangle=\d_{kk'}\Theta\left(-\cos\left(\frac{\pi k}{L+1}\right)\right)\,,
\ee
where $\Theta(x)$ is the Heaviside step function. The initial fermionic two-point correlation function then can be obtained as
\be 
C_{mn}(0)=\langle c_m^{\dg}c_n\rangle=\frac{2}{L+1}\sum_{k=1}^{L}\sin\left(\frac{\pi km}{L+1}\right)\sin\left(\frac{\pi kn}{L+1}\right)\Theta\left(-\cos\left(\frac{\pi k}{L+1}\right)\right)\,.
\ee
\par At $t=0$, the system is suddenly changed to evolve according to a new non-uniform Hamiltonian (i.e. the M\"obius/SSD Hamiltonian)
\be\label{H1lat}
H_1(\th)=\frac12\sum_{n=1}^{L-1}\left(1-\tanh(2\th)\cos(\frac{\pi(n+1/2)}{L})\right)c_n^{\dg}c_{n+1}+h.c.\,.
\ee
To derive the time evolution of the fermionic two-point correlation function, we write $H_1(\th)$ as 
\be
H_1(\th)=\sum_{m,n}h_{mn}c_m^{\dg}c_n\,,
\ee
where we have introduced a $L\times L$ matrix $h$ with elements
\be 
h_{mn}=\frac12\left(1-\tanh(2\th)\cos(\frac{\pi(n+1/2)}{L})\right)(\d_{m,n+1}+\d_{m,n-1})\,.
\ee
It's straightforward to prove that\footnote{In \cite{Caceffo:2024qce}, the authors obtained quite general formulas for the non-equilibrium fermionic two-point correlation functions in the presence of gain/loss dissipation} (see appendix \ref{appB} for details)
\be 
C(t)=e^{iht}\cdot C(0)\cdot e^{-iht}\,,
\ee
where $e^{iht}$ and $C(0)$ are all $L\times L$ matrices and we use $\cdot$ to represent matrix multiplication. From the $L\times L$ matrix $C(t)$, we obtain a $l\times l$ matrix $C_A(t)$ by restricting the coulumn and row index within the subsystem $A$. Denoting the spectrum of the matrix $C_A(t)$ as $\{\l_j(t)\}$, the charged moments are obtained from
\be 
Z_n(\a,t)=\prod_{j=1}^l[e^{i\a/2}\l_j(t)^n+e^{-i\a/2}(1-\l_j(t))^n]\,.
\ee
To numerically compute the SREE, we rewrite Eq.~(\ref{Sq}) as
\be\label{Sqn} 
S_A(q)=\log\mZ_1(q)-\frac{1}{\mZ_1(q)}\int_{-\pi}^{\pi}\frac{d\a}{2\pi}e^{-i\a q}Z_n(\a)\p_n\log Z_n(\a)|_{n=1}\,.
\ee
Here and in the following discussions, for simplicity, we omit the variable $t$. In Eq.~(\ref{Sqn}), $\mZ_1(q)$ is simply computed from $Z_1(\a)$ via Fourier transformation. The most complicated part is the integrand, and the only additional formula we need is
\be 
\p_n\log Z_n(\a)|_{n=1}=\sum_{j=1}^l\frac{e^{i\a}\l_j\log\l_j+(1-\l_j)\log(1-\l_j)}{e^{i\a}\l_j+(1-\l_j)}\,.
\ee
\par The numerical data of the charged moment $Z_1(\a,t)$ and the SREE $S_A(q,t)$ is shown in figure \ref{fig1}. As shown in the figure, the match between numerical computation and analytical method are perfect.  \footnote{For scenarios where $q\geq 2$, a much larger $L$ must be employed to achieve satisfactory alignment, though this requires computational resources that exceed the capacity of personal computing systems.}
\section{Conclusion}
In this paper, we investigate the dynamics of symmetry-resolved entanglement entropy (SREE) in a finite system following an inhomogeneous quench—specifically, the Möbius/SSD quench—using both analytical and numerical approaches. Our analytical treatment employs techniques from 2D CFT, leveraging the solvability of the Möbius deformation. Meanwhile, our exact numerical method is broadly applicable to SREE computations in other inhomogeneous quench problems.
\par Key findings include:
\par (a)~Equipartition of entanglement: At leading order in the large-$L$ limit, the entanglement entropy exhibits equipartition among symmetry sectors, while subleading corrections break this behavior.
\par (b)~Unbounded logarithmic growth: At late times, the SREE grows as 
$\log t$ without saturation, indicating the suppression of boundary effects.
\par These results suggest intriguing future directions, such as studying SREE dynamics in other quench protocols within the Möbius deformation framework \cite{Goto:2021sqx, Nozaki:2023fkx}. 
\section*{Acknowledgments}
We would like to thank Jun-Bao Wu and Pasquale Calabrese for very helpful discussions. H. C. was supported by the National Natural Science Foundation of China (Grant Nos.12005081, and 12465014), and Program of China Scholarship Council (Grant No. 202308360098). M. Z. was supported by the National Natural Science Foundation of China (Grant No. 12365010) and Program of Chinese Scholarship Council Scholarship (Grant No. 202208360067).
\begin{appendix}
\section{Details of conformal mappings}
In this section, we give the details of the conformal mapping method in deriving the time evolution of the charged moments in section \ref{sec4.2}.
\par At $w=w_0=il$, a direct computation leads to 
\be 
\frac{\p z}{\p w}\frac{\p\bar z}{\p\bar w}\Big|_{w=w_0}=\frac{4\pi^2}{L^2}\,,
\ee
and the Jacobian of the mapping from $z$ to $z'$ is 
\be 
\begin{split}
\frac{\p z'}{\p z}\frac{\p\bar z'}{\p\bar z}\Big|_{w=w_0}=&16\l^2[(\l-1)^2\cosh(4\theta)-(2\l^2-2)\cosh(2\theta)\\
&+(\lambda +1)^2+(2\l-2)\sinh(2\theta)((1-\l)\cosh(2\theta )+\l+1)\cos(2\pi l/L)]^{-2}\,.
\end{split}
\ee
At the SSD point, which corresponds to taking $\th\to\infty$, we have
\be 
\frac{\p z'}{\p z}\frac{\p\bar z'}{\p\bar z}\Big|_{w=w_0}=\frac{L^2}{4(\pi^2\tau^2-\pi\tau L)\sin^2(2\pi l/L)+L^2}\,.
\ee
At $w=w_0=il$, the new variable $z'$ defined in Eq.~(\ref{zp}) becomes
\be 
z'|_{w=w_0}=\frac{(1-\l)\sinh(2\theta)+e^{\frac{2i\pi l}{L}}((\l-1)\cosh(2\theta)+\l +1)}{(1-\l)\cosh(2\theta )+\l +(\l-1)\sinh(2\theta) e^{\frac{2i\pi l}{L}}+1}\,.
\ee
In the SSD case, this expression simplifies to
\be 
z'|_{w=w_0}=\frac{e^{\frac{2i\pi l}{L}}(L+\pi\tau)-\pi\tau}{L+\pi\tau\left(e^{\frac{2i\pi l}{L}}-1\right)}\,.
\ee
\section{Time evolution of fermionic correlation function}\label{appB}
To derive the time evolution of the fermionic two-point correlation function for the given Hamiltonian $H = \sum_{m,n} h_{mn} c_m^\dg c_n$. Here $h$ is a real matrix. The hermiticity of the Hamiltonian further requires $h=h^{T}$. We start by considering the Heisenberg equations of motion for the fermionic operators $c_m^\dg(t)$ and $c_n(t)$. For $c_m^\dg(t)$, we have
\be
\frac{d}{dt} c_m^\dg(t) = i[H, c_m^\dg(t)] = i\sum_p h_{pm} c_p^\dg(t)\,,
\ee
and similar equation hold for $c_n(t)$ as
\be
\frac{d}{dt} c_n(t) = i [H, c_n(t)] = -i\sum_q h_{nq} c_q(t)\,.
\ee
The fermionic two-point correlation function is $C_{mn}(t) = \langle c_m^\dg(t) c_n(t) \rangle$. Using the product rule, its time derivative is
\be
\frac{d}{dt} C_{mn}(t) = \left\langle\frac{d c_m^\dg(t)}{dt} c_n(t) \right\rangle + \left\langle c_m^\dg(t) \frac{d c_n(t)}{dt} \right\rangle\,.
\ee
Substituting the Heisenberg equations of motion, we have
\be
\frac{d}{dt} C_{mn}(t) = i\sum_p h_{pm}\langle c_p^\dg(t)c_n(t)\rangle-i \sum_q h_{nq} \langle c_m^\dg(t) c_q(t)\rangle\,.
\ee
Recognizing that $\langle c_p^\dg c_n \rangle = C_{pn}$ and $\langle c_m^\dg c_q \rangle = C_{mq}$, we rewrite the time derivative as
\be
\frac{d}{dt} C_{mn}(t)=i\sum_p (h_{pm}C_{pn}(t)-h_{nq}C_{mq}(t))\,.
\ee
This can be expressed in matrix form as
\be
\frac{d}{dt} C(t) = i [h, C(t)]\,,
\ee
where we have used the fact that $h$ is a symmetric matrix. The solution of this differential equation is given by
\be
C(t) = e^{i h t}\cdot C(0)\cdot e^{-i h t}\,.
\ee

\end{appendix}

\bibliographystyle{JHEP}
\bibliography{biblio.bib}

\providecommand{\href}[2]{#2}\begingroup\raggedright\begin{thebibliography}{10}

\bibitem{nahum2017quantum}
A.~Nahum, J.~Ruhman, S.~Vijay and J.~Haah, \emph{Quantum entanglement growth
  under random unitary dynamics}, {\emph{Physical Review X} {\bfseries 7}
  (2017) 031016}.

\bibitem{zhou2020entanglement}
T.~Zhou and A.~Nahum, \emph{Entanglement membrane in chaotic many-body
  systems}, {\emph{Physical Review X} {\bfseries 10} (2020) 031066}.

\bibitem{Ryu:2006bv}
S.~Ryu and T.~Takayanagi, \emph{{Holographic derivation of entanglement entropy
  from AdS/CFT}},
  \href{https://doi.org/10.1103/PhysRevLett.96.181602}{\emph{Phys. Rev. Lett.}
  {\bfseries 96} (2006) 181602}
  [\href{https://arxiv.org/abs/hep-th/0603001}{{\ttfamily hep-th/0603001}}].

\bibitem{Takayanagi:2010wp}
T.~Takayanagi and T.~Ugajin, \emph{{Measuring Black Hole Formations by
  Entanglement Entropy via Coarse-Graining}},
  \href{https://doi.org/10.1007/JHEP11(2010)054}{\emph{JHEP} {\bfseries 11}
  (2010) 054} [\href{https://arxiv.org/abs/1008.3439}{{\ttfamily 1008.3439}}].

\bibitem{Almheiri:2020cfm}
A.~Almheiri, T.~Hartman, J.~Maldacena, E.~Shaghoulian and A.~Tajdini,
  \emph{{The entropy of Hawking radiation}},
  \href{https://doi.org/10.1103/RevModPhys.93.035002}{\emph{Rev. Mod. Phys.}
  {\bfseries 93} (2021) 035002}
  [\href{https://arxiv.org/abs/2006.06872}{{\ttfamily 2006.06872}}].

\bibitem{Chen:2020hmv}
H.Z.~Chen, R.C.~Myers, D.~Neuenfeld, I.A.~Reyes and J.~Sandor, \emph{{Quantum
  Extremal Islands Made Easy, Part II: Black Holes on the Brane}},
  \href{https://doi.org/10.1007/JHEP12(2020)025}{\emph{JHEP} {\bfseries 12}
  (2020) 025} [\href{https://arxiv.org/abs/2010.00018}{{\ttfamily
  2010.00018}}].

\bibitem{Grimaldi:2022suv}
G.~Grimaldi, J.~Hernandez and R.C.~Myers, \emph{{Quantum extremal islands made
  easy. Part IV. Massive black holes on the brane}},
  \href{https://doi.org/10.1007/JHEP03(2022)136}{\emph{JHEP} {\bfseries 03}
  (2022) 136} [\href{https://arxiv.org/abs/2202.00679}{{\ttfamily
  2202.00679}}].

\bibitem{2009PThPh.122..953G}
A.~{Gendiar}, R.~{Krcmar} and T.~{Nishino}, \emph{{Spherical Deformation for
  One-Dimensional Quantum Systems}},
  \href{https://doi.org/10.1143/PTP.122.953}{\emph{Progress of Theoretical
  Physics} {\bfseries 122} (2009) 953}
  [\href{https://arxiv.org/abs/0810.0622}{{\ttfamily 0810.0622}}].

\bibitem{2009PThPh.122..953Gerr}
A.~{Gendiar}, R.~{Krcmar} and T.~{Nishino}, \emph{{Spherical Deformation for
  One-Dimensional Quantum Systems}},
  \href{https://doi.org/10.1143/PTP.122.953}{\emph{Progress of Theoretical
  Physics} {\bfseries 123} (2009) 393}.

\bibitem{Hikihara:2011mtb}
T.~Hikihara and T.~Nishino, \emph{{Connecting distant ends of one-dimensional
  critical systems by a sine-square deformation}},
  \href{https://doi.org/10.1103/physrevb.83.060414}{\emph{Phys. Rev. B}
  {\bfseries 83} (2011) 060414}.

\bibitem{Katsura:2011ss}
H.~Katsura, \emph{{Sine-square deformation of solvable spin chains and
  conformal field theories}},
  \href{https://doi.org/10.1088/1751-8113/45/11/115003}{\emph{J. Phys. A}
  {\bfseries 45} (2012) 115003}
  [\href{https://arxiv.org/abs/1110.2459}{{\ttfamily 1110.2459}}].

\bibitem{2011PhRvA..83e2118G}
A.~{Gendiar}, M.~{Dani{\v{s}}ka}, Y.~{Lee} and T.~{Nishino}, \emph{{Suppression
  of finite-size effects in one-dimensional correlated systems}},
  \href{https://doi.org/10.1103/PhysRevA.83.052118}{\emph{pra} {\bfseries 83}
  (2011) 052118} [\href{https://arxiv.org/abs/1012.1472}{{\ttfamily
  1012.1472}}].

\bibitem{Shibata:2011jup}
N.~Shibata and C.~Hotta, \emph{{Boundary effects in the density-matrix
  renormalization group calculation}},
  \href{https://doi.org/10.1103/physrevb.84.115116}{\emph{Phys. Rev. B}
  {\bfseries 84} (2011) 115116}.

\bibitem{Maruyama:2011njv}
I.~Maruyama, H.~Katsura and T.~Hikihara, \emph{{Sine-square deformation of free
  fermion systems in one and higher dimensions}},
  \href{https://arxiv.org/abs/1108.2973}{{\ttfamily 1108.2973}}.

\bibitem{Katsura:2011zyx}
H.~Katsura, \emph{{Exact ground state of the sine-square deformed XY spin
  chain}}, \href{https://doi.org/10.1088/1751-8113/44/25/252001}{\emph{J. Phys.
  A} {\bfseries 44} (2011) 252001}
  [\href{https://arxiv.org/abs/1104.1721}{{\ttfamily 1104.1721}}].

\bibitem{Okunishi:2016zat}
K.~Okunishi, \emph{{Sine-square deformation and M\"obius quantization of 2D
  conformal field theory}},
  \href{https://doi.org/10.1093/ptep/ptw060}{\emph{PTEP} {\bfseries 2016}
  (2016) 063A02} [\href{https://arxiv.org/abs/1603.09543}{{\ttfamily
  1603.09543}}].

\bibitem{Wen:2016inm}
X.~Wen, S.~Ryu and A.W.W.~Ludwig, \emph{{Evolution operators in conformal field
  theories and conformal mappings: Entanglement Hamiltonian, the sine-square
  deformation, and others}},
  \href{https://doi.org/10.1103/PhysRevB.93.235119}{\emph{Phys. Rev. B}
  {\bfseries 93} (2016) 235119}
  [\href{https://arxiv.org/abs/1604.01085}{{\ttfamily 1604.01085}}].

\bibitem{Goto:2021sqx}
K.~Goto, M.~Nozaki, K.~Tamaoka, M.T.~Tan and S.~Ryu, \emph{{Non-Equilibrating a
  Black Hole with Inhomogeneous Quantum Quench}},
  \href{https://arxiv.org/abs/2112.14388}{{\ttfamily 2112.14388}}.

\bibitem{Goto:2023wai}
K.~Goto, M.~Nozaki, S.~Ryu, K.~Tamaoka and M.T.~Tan, \emph{{Scrambling and
  recovery of quantum information in inhomogeneous quenches in two-dimensional
  conformal field theories}},
  \href{https://doi.org/10.1103/PhysRevResearch.6.023001}{\emph{Phys. Rev.
  Res.} {\bfseries 6} (2024) 023001}
  [\href{https://arxiv.org/abs/2302.08009}{{\ttfamily 2302.08009}}].

\bibitem{Nozaki:2023fkx}
M.~Nozaki, K.~Tamaoka and M.T.~Tan, \emph{{Inhomogeneous quenches as state
  preparation in two-dimensional conformal field theories}},
  \href{https://doi.org/10.1103/PhysRevD.109.126014}{\emph{Phys. Rev. D}
  {\bfseries 109} (2024) 126014}
  [\href{https://arxiv.org/abs/2310.19376}{{\ttfamily 2310.19376}}].

\bibitem{Wen:2018vux}
X.~Wen and J.-Q.~Wu, \emph{{Quantum dynamics in sine-square deformed conformal
  field theory: Quench from uniform to nonuniform conformal field theory}},
  \href{https://doi.org/10.1103/PhysRevB.97.184309}{\emph{Phys. Rev. B}
  {\bfseries 97} (2018) 184309}
  [\href{https://arxiv.org/abs/1802.07765}{{\ttfamily 1802.07765}}].

\bibitem{Goto:2023yxb}
K.~Goto, T.~Guo, T.~Nosaka, M.~Nozaki, S.~Ryu and K.~Tamaoka, \emph{{Spatial
  deformation of many-body quantum chaotic systems and quantum information
  scrambling}}, \href{https://doi.org/10.1103/PhysRevB.109.054301}{\emph{Phys.
  Rev. B} {\bfseries 109} (2024) 054301}
  [\href{https://arxiv.org/abs/2305.01019}{{\ttfamily 2305.01019}}].

\bibitem{maccormack2019holographic}
I.~MacCormack, A.~Liu, M.~Nozaki and S.~Ryu, \emph{Holographic duals of
  inhomogeneous systems: the rainbow chain and the sine-square deformation
  model}, {\emph{Journal of Physics A: Mathematical and Theoretical} {\bfseries
  52} (2019) 505401}.

\bibitem{Bernamonti:2024fgx}
A.~Bernamonti, F.~Galli and D.~Ge, \emph{{Boundary-induced transitions in
  M\"obius quenches of holographic BCFT}},
  \href{https://doi.org/10.1007/JHEP06(2024)184}{\emph{JHEP} {\bfseries 06}
  (2024) 184} [\href{https://arxiv.org/abs/2402.16555}{{\ttfamily
  2402.16555}}].

\bibitem{Goldstein:2017bua}
M.~Goldstein and E.~Sela, \emph{{Symmetry-resolved entanglement in many-body
  systems}}, \href{https://doi.org/10.1103/PhysRevLett.120.200602}{\emph{Phys.
  Rev. Lett.} {\bfseries 120} (2018) 200602}
  [\href{https://arxiv.org/abs/1711.09418}{{\ttfamily 1711.09418}}].

\bibitem{Murciano:2020vgh}
S.~Murciano, G.~Di~Giulio and P.~Calabrese, \emph{{Entanglement and symmetry
  resolution in two dimensional free quantum field theories}},
  \href{https://doi.org/10.1007/JHEP08(2020)073}{\emph{JHEP} {\bfseries 08}
  (2020) 073} [\href{https://arxiv.org/abs/2006.09069}{{\ttfamily
  2006.09069}}].

\bibitem{Chen:2021pls}
H.-H.~Chen, \emph{{Symmetry decomposition of relative entropies in conformal
  field theory}}, \href{https://doi.org/10.1007/JHEP07(2021)084}{\emph{JHEP}
  {\bfseries 07} (2021) 084}
  [\href{https://arxiv.org/abs/2104.03102}{{\ttfamily 2104.03102}}].

\bibitem{Capizzi:2021zga}
L.~Capizzi and P.~Calabrese, \emph{{Symmetry resolved relative entropies and
  distances in conformal field theory}},
  \href{https://doi.org/10.1007/JHEP10(2021)195}{\emph{JHEP} {\bfseries 10}
  (2021) 195} [\href{https://arxiv.org/abs/2105.08596}{{\ttfamily
  2105.08596}}].

\bibitem{Chen:2021nma}
H.-H.~Chen, \emph{{Charged R\'enyi negativity of massless free bosons}},
  \href{https://doi.org/10.1007/JHEP02(2022)117}{\emph{JHEP} {\bfseries 02}
  (2022) 117} [\href{https://arxiv.org/abs/2111.11028}{{\ttfamily
  2111.11028}}].

\bibitem{Bertini:2022srv}
B.~Bertini, P.~Calabrese, M.~Collura, K.~Klobas and C.~Rylands,
  \emph{{Nonequilibrium Full Counting Statistics and Symmetry-Resolved
  Entanglement from Space-Time Duality}},
  \href{https://doi.org/10.1103/PhysRevLett.131.140401}{\emph{Phys. Rev. Lett.}
  {\bfseries 131} (2023) 140401}
  [\href{https://arxiv.org/abs/2212.06188}{{\ttfamily 2212.06188}}].

\bibitem{Northe:2023khz}
C.~Northe, \emph{{Entanglement Resolution with Respect to Conformal Symmetry}},
  \href{https://doi.org/10.1103/PhysRevLett.131.151601}{\emph{Phys. Rev. Lett.}
  {\bfseries 131} (2023) 151601}
  [\href{https://arxiv.org/abs/2303.07724}{{\ttfamily 2303.07724}}].

\bibitem{Li:2023zgy}
P.~Li and Y.~Ling, \emph{{Refined symmetry-resolved Page curve and charged
  black holes*}}, \href{https://doi.org/10.1088/1674-1137/ad2e83}{\emph{Chin.
  Phys. C} {\bfseries 48} (2024) 053109}
  [\href{https://arxiv.org/abs/2311.04436}{{\ttfamily 2311.04436}}].

\bibitem{Banerjee:2024ldl}
A.~Banerjee, R.~Basu, A.~Bhattacharyya and N.~Chakrabarti, \emph{{Symmetry
  resolution in non-Lorentzian field theories}},
  \href{https://doi.org/10.1007/JHEP06(2024)121}{\emph{JHEP} {\bfseries 06}
  (2024) 121} [\href{https://arxiv.org/abs/2404.02206}{{\ttfamily
  2404.02206}}].

\bibitem{Yan:2024rcl}
F.~Yan, S.~Murciano, P.~Calabrese and R.~Konik, \emph{{On symmetry-resolved
  generalized entropies}},  \href{https://arxiv.org/abs/2412.14165}{{\ttfamily
  2412.14165}}.

\bibitem{Huang:2025lsy}
Y.~Huang and Y.~Zhou, \emph{{Symmetry-Resolved Entanglement Entropy in Higher
  Dimensions}},  \href{https://arxiv.org/abs/2503.09070}{{\ttfamily
  2503.09070}}.

\bibitem{Feldman:2019upn}
N.~Feldman and M.~Goldstein, \emph{{Dynamics of Charge-Resolved Entanglement
  after a Local Quench}},
  \href{https://doi.org/10.1103/PhysRevB.100.235146}{\emph{Phys. Rev. B}
  {\bfseries 100} (2019) 235146}
  [\href{https://arxiv.org/abs/1905.10749}{{\ttfamily 1905.10749}}].

\bibitem{Parez:2021pgq}
G.~Parez, G.~Parez, R.~Bonsignori, R.~Bonsignori, P.~Calabrese and
  P.~Calabrese, \emph{{Exact quench dynamics of symmetry resolved entanglement
  in a free fermion chain}},
  \href{https://doi.org/10.1088/1742-5468/ac21d7}{\emph{J. Stat. Mech.}
  {\bfseries 2109} (2021) 093102}
  [\href{https://arxiv.org/abs/2106.13115}{{\ttfamily 2106.13115}}].

\bibitem{Chen:2022gyy}
H.-H.~Chen, \emph{{Dynamics of charge imbalance resolved negativity after a
  global quench in free scalar field theory}},
  \href{https://doi.org/10.1007/JHEP08(2022)146}{\emph{JHEP} {\bfseries 08}
  (2022) 146} [\href{https://arxiv.org/abs/2205.09532}{{\ttfamily
  2205.09532}}].

\bibitem{Oshima:2022yrw}
H.~Oshima and Y.~Fuji, \emph{{Charge fluctuation and charge-resolved
  entanglement in a monitored quantum circuit with U(1) symmetry}},
  \href{https://doi.org/10.1103/PhysRevB.107.014308}{\emph{Phys. Rev. B}
  {\bfseries 107} (2023) 014308}
  [\href{https://arxiv.org/abs/2210.16009}{{\ttfamily 2210.16009}}].

\bibitem{Scopa:2022gfw}
S.~Scopa and D.X.~Horv\'ath, \emph{{Exact hydrodynamic description of
  symmetry-resolved R\'enyi entropies after a quantum quench}},
  \href{https://doi.org/10.1088/1742-5468/ac85eb}{\emph{J. Stat. Mech.}
  {\bfseries 2208} (2022) 083104}
  [\href{https://arxiv.org/abs/2205.02924}{{\ttfamily 2205.02924}}].

\bibitem{Murciano:2023zvk}
S.~Murciano, P.~Calabrese and V.~Alba, \emph{{Symmetry-resolved entanglement in
  fermionic systems with dissipation}},
  \href{https://doi.org/10.1088/1742-5468/ad0224}{\emph{J. Stat. Mech.}
  {\bfseries 2311} (2023) 113102}
  [\href{https://arxiv.org/abs/2303.12120}{{\ttfamily 2303.12120}}].

\bibitem{Caceffo:2024jbc}
F.~Caceffo, S.~Murciano and V.~Alba, \emph{{Entangled multiplets, asymmetry,
  and quantum Mpemba effect in dissipative systems}},
  \href{https://doi.org/10.1088/1742-5468/ad4537}{\emph{J. Stat. Mech.}
  {\bfseries 2024} (2024) 063103}
  [\href{https://arxiv.org/abs/2402.02918}{{\ttfamily 2402.02918}}].

\bibitem{Chen:2023whs}
H.-H.~Chen and Z.-X.~Huang, \emph{{Dynamics of charge imbalance resolved
  negativity after a local joining quench}},
  \href{https://doi.org/10.1007/JHEP12(2023)128}{\emph{JHEP} {\bfseries 12}
  (2023) 128} [\href{https://arxiv.org/abs/2308.02868}{{\ttfamily
  2308.02868}}].

\bibitem{Castro-Alvaredo:2024azg}
O.A.~Castro-Alvaredo and L.~Santamar\'\i{}a-Sanz, \emph{{Symmetry-resolved
  measures in quantum field theory: A short review}},
  \href{https://doi.org/10.1142/S0217984924300023}{\emph{Mod. Phys. Lett. B}
  {\bfseries 39} (2025) 2430002}
  [\href{https://arxiv.org/abs/2403.06652}{{\ttfamily 2403.06652}}].

\bibitem{Das:2017sgp}
D.~Das, S.R.~Das, D.A.~Galante, R.C.~Myers and K.~Sengupta, \emph{{An exactly
  solvable quench protocol for integrable spin models}},
  \href{https://doi.org/10.1007/JHEP11(2017)157}{\emph{JHEP} {\bfseries 11}
  (2017) 157} [\href{https://arxiv.org/abs/1706.02322}{{\ttfamily
  1706.02322}}].

\bibitem{Das:2014jna}
S.R.~Das, D.A.~Galante and R.C.~Myers, \emph{{Universal scaling in fast quantum
  quenches in conformal field theories}},
  \href{https://doi.org/10.1103/PhysRevLett.112.171601}{\emph{Phys. Rev. Lett.}
  {\bfseries 112} (2014) 171601}
  [\href{https://arxiv.org/abs/1401.0560}{{\ttfamily 1401.0560}}].

\bibitem{Das:2014hqa}
S.R.~Das, D.A.~Galante and R.C.~Myers, \emph{{Universality in fast quantum
  quenches}}, \href{https://doi.org/10.1007/JHEP02(2015)167}{\emph{JHEP}
  {\bfseries 02} (2015) 167} [\href{https://arxiv.org/abs/1411.7710}{{\ttfamily
  1411.7710}}].

\bibitem{Myers:2017sxr}
R.C.~Myers, M.~Rozali and B.~Way, \emph{{Holographic Quenches in a Confined
  Phase}}, \href{https://doi.org/10.1088/1751-8121/aa927c}{\emph{J. Phys. A}
  {\bfseries 50} (2017) 494002}
  [\href{https://arxiv.org/abs/1706.02438}{{\ttfamily 1706.02438}}].

\bibitem{calabrese2016quantum}
P.~Calabrese and J.~Cardy, \emph{Quantum quenches in 1+ 1 dimensional conformal
  field theories}, {\emph{Journal of Statistical Mechanics: Theory and
  Experiment} {\bfseries 2016} (2016) 064003}.

\bibitem{Bonsignori:2020laa}
R.~Bonsignori and P.~Calabrese, \emph{{Boundary effects on symmetry resolved
  entanglement}}, \href{https://doi.org/10.1088/1751-8121/abcc3a}{\emph{J.
  Phys. A} {\bfseries 54} (2021) 015005}
  [\href{https://arxiv.org/abs/2009.08508}{{\ttfamily 2009.08508}}].

\bibitem{Bonsignori:2019naz}
R.~Bonsignori, P.~Ruggiero and P.~Calabrese, \emph{{Symmetry resolved
  entanglement in free fermionic systems}},
  \href{https://doi.org/10.1088/1751-8121/ab4b77}{\emph{J. Phys. A} {\bfseries
  52} (2019) 475302} [\href{https://arxiv.org/abs/1907.02084}{{\ttfamily
  1907.02084}}].

\bibitem{jin2004quantum}
B.-Q.~Jin and V.E.~Korepin, \emph{Quantum spin chain, toeplitz determinants and
  the fisher—hartwig conjecture}, {\emph{Journal of statistical physics}
  {\bfseries 116} (2004) 79}.

\bibitem{Xavier:2018kqb}
J.C.~Xavier, F.C.~Alcaraz and G.~Sierra, \emph{{Equipartition of the
  entanglement entropy}},
  \href{https://doi.org/10.1103/PhysRevB.98.041106}{\emph{Phys. Rev. B}
  {\bfseries 98} (2018) 041106}
  [\href{https://arxiv.org/abs/1804.06357}{{\ttfamily 1804.06357}}].

\bibitem{Caceffo:2024qce}
F.~Caceffo and V.~Alba, \emph{{Fate of entanglement in quadratic Markovian
  dissipative systems}},  \href{https://arxiv.org/abs/2406.15328}{{\ttfamily
  2406.15328}}.

\end{thebibliography}\endgroup

\end{document}